# A Cyber Physical System Framework for UAV Communications

Haijun Wang, Haitao Zhao, Dongtang Ma, and Jibo Wei

*Abstract*—Diverse applications have witnessed the prevalence of unmanned aerial vehicles (UAVs) due to their agility and versatility. Compared with computation and control, the communication tends to be the bottleneck of the whole UAV system. Cyber physical system (CPS), which achieves the integration of the cyber and physical domains, can inspire us to deal with the communication problems through a cross-disciplinary method. To this end, we first expound the coupling effects of computation and control to communication. Then, we propose a novel CPS framework for UAV communications. By extending the dimension of communication decisions to computation and control, the framework can precisely orient and settle the communication issues. Further, a quantitative energy optimization model is established to guide the protocol and algorithm design for UAV communications. Case simulation results validate the CPS framework in terms of the energy consumption and communication delay.

## INTRODUCTION

Along with the growing popularization of unmanned aerial vehicles (UAVs) in diverse areas, the communication problem is getting prominent. Compared with computation and control, UAV communications have much more preconditions, including spectrum and channel, except for the common energy. It is also more vulnerable to external factors, including electromagnetic interference and geographical/channel variation, especially under harsh environments (e.g., disaster relief and military operations) [1]. Furthermore, intermittent links, fluid topologies and Doppler effect resulted from the intrinsic mobility of UAVs will make these challenges more intractable [2]. Therefore, the communication of UAVs tends to be the bottleneck of the whole system, which needs paramount attentions.

Cognitive radio (CR) technology has emerged as a promising solution to alleviate the spectrum scarcity problem of UAV communications. Its dynamic spectrum access technique can handle with spectrum variations and outside interferences. From some extent, CR has propelled UAV communications into a smart era. Nevertheless, CR is usually confined to the communication dimension. Therefore, it is incompetent in settling some communication issues that are unique to UAVs. With traditional chip giants advancing into the UAV market, many high-performance computation suites are released for drones. Additionally, open source software projects on artificial intelligence (AI) are springing up, which can be applied to UAVs. All these significantly promote the computation power, to be exact, the intelligence of UAVs. Correspondingly, the flight control performance of UAVs is also boosted. In this regard, it is asserted that UAVs are becoming more and more intelligent.

There is a trend that communication, computation and control modules are integrally scheduled for intelligent UAV networks [3]. Since computation and control are much more powerful than communication comparatively, it is possible to transfer the surplus power of the former two into the communication ability. In other words, they should be well tuned for better system performance under the shared and limited resources. This is rational considering the three cyber issues (i.e., communication, computation and control) will influence and gain from each other in a coupling manner. For instance, when communication links suffers shadow effects, UAVs can adjust their locations for pursuing line of sight channels via flight control, instead of only modifying communication parameters. Therefore, a cross-disciplinary outlook may be enlightening to cope with the problems in UAV communications.

As an extension of embedded system with communication modules, computation abilities and control devices, UAV can be modeled as a cyber physical system [4]. It achieves the implantation of the cyber issues into physical devices, which can be enlightening to deal with communication problems through a cross-disciplinary method. As the next generation of systems, CPS is envisaged as a critical technique to achieve AI [5]. Consequently, incorporating a CPS viewpoint into UAV systems is promising to remarkably boost the communication performance.

Existing researches, indeed few, just simply analyze the coupling effects among the cyber and physical aspects, or investigate design issues and challenges on UAV communications [4, 6]. Although [7] provides a unified framework to specify the quantitative contribution of each cyber and physical aspect to the system performance, its focus is not on settling communication issues. In a word, there lacks a unified CPS framework and a quantitative analysis model for guiding the specific protocol/algorithm design of UAV communications.

In this article, inspired by the coupling effects among the cyber and physical issues, we propose a novel CPS framework for UAV communications. It incorporates computation and control issues to extend the dimensions of communication decisions. Further, a quantitative energy optimization model is established which is expected to instruct the protocol and algorithm design for UAV communications.

## COUPLING EFFECTS IN UAV COMMUNICATIONS

Computation and control will pose positive effects on communication. Revealing these coupling effects can deeply enlighten us to address communication issues with a cross-disciplinary method.

### COMPUTATION TO COMMUNICATION

Computation can promote communication capability. It may contribute to reach the communication capacity bound, that is, Shannon's law [8].

With the prosperity of AI, communication is undergoing a

*The authors are with the National University of Defense Technology.*

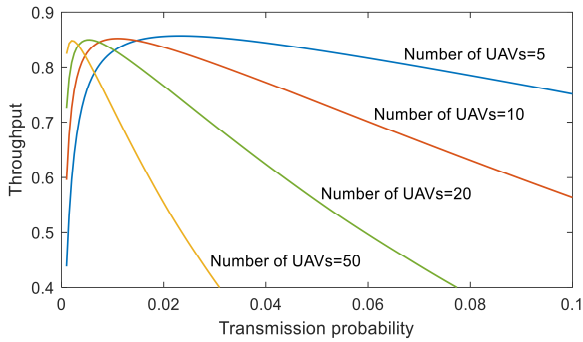

Figure 1. Throughput versus transmission probability for the DCF protocol under the basic access mechanism.

revolution to achieve intelligence. Nevertheless, how to embed intelligence into communication is still an open topic.

Intuitively, computation can help to obtain the optimal communication parameters according to circumstance fluctuations. One method that may implement intelligent communication can be proposed as follows. When a communication demand emerges, what computation will do is, first, modelling the channel by extracting the geographical and spectrum characteristics; then, categorizing the channel model based on predefined rules; and lastly, matching and loading the optimal communication configurations from off-the-shelf waveform/algorithm/protocol libraries, or directly making fresh communication decisions on the basis of the measurements. For UAVs with learning capability, the more this process repeats, the better they will communicate. It means that UAVs can evolve intelligently to accommodate unfamiliar circumstances.

From another perspective, on the contrary with traditional communications which target for high throughput, computation can be exploited to reduce communication volume and overhead of the whole system. The nature of communication is to efficiently transfer valuable information, not just bulky data block. To this end, computation may contribute a lot at least from two levels. At individual node level, data redundancy can be eliminated by preprocessing the captured data to extract the key information. For example, in a reconnaissance mission, the target and its detailed motion pattern can be obtained through pattern recognition techniques. In this way, the UAV only needs to transmit a concise text (e.g., "a red car is moving to the north with the speed of 60 km/h") rather than high-quality videos. At network level, information delivery which will not bring evident improvement to the task performance can be cut down. To this end, multiagent decision technique can be adopted to capacitate UAVs decide whether/what/whom/when to communicate [9]. The most valuable information can be transmitted to the most wanted nodes at the most proper moment. Therefore, computation will make communication more accurate. In return, communication resources could be economized, and the performance of essential communications can be enhanced. Figure 1 shows the theoretically achievable maximum throughput by the Distributed Coordination Function (DCF) protocol in the case of basic access mechanism. The simulation parameters are the same as that in [10]. It indicates that decreasing the transmission probability of individuals (e.g., lessening data volume from the source) instead augments the throughput in some cases. In that sense, to reduce communication volume through computation happens to coincide with the target of traditional communications.

*CONTROL TO COMMUNICATION*

As the inherent attribute of UAVs, mobility is dominated by flight control. It will cause various challenges to UAV communications and networking. Conversely, flight control extends the dimension of communication decisions by assimilating physical issues, for example, changing locations. As two primary energy consumers, communication and control should be orchestrated well.

The communication network is generally insured by the optimal formation control, or at least, the appropriate distance maintenance among UAVs. Long distance would call for higher transmission power. This will be alleviated through diminishing the distance, that is, one moving to access the other. It can lessen the transmission power while reaping equivalent communication performance. Admittedly, both magnifying the transmission power and shifting the UAV will consume additional energy. Actually, augmenting the transmission power may be unreasonable because higher transmission power implies stronger interference to others and lower spectrum reusage. Sometimes, it may be completely unavailing particularly when there exist strong shadow effects. Therefore, flight control would be better. Figure 2 illustrates three typical scenarios of controlling the mobility to optimize UAV communications, which are discussed as follows.

*i) Eliminate channel fading*: UAVs fly to pursue high-quality channel especially when there exists strong shadow fading. In this case, the optimal location for transmission and the task direction should be jointly considered.

*ii) Store-carry-forward*: Depleting UAVs will carry and deliver data to the desiring UAVs when they return for charging. Therefore, the throughput and access efficiency can be improved [11]. Besides, energetic UAVs can fly back and forth to relay data for two remote areas [12]. The difference lies in that the latter proactively converts the mobility energy into the communication performance.

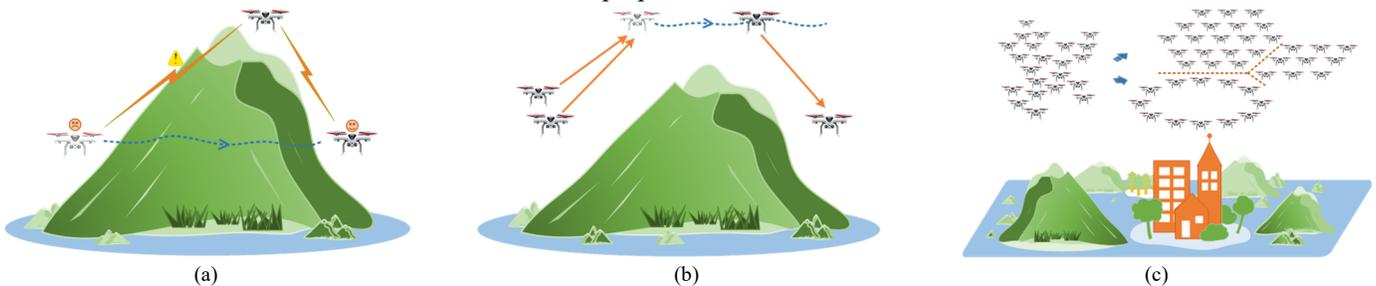

Figure 2. Three typical scenarios of controlling mobility to optimize communications: a) eliminate channel fading; b) store-carry-forward; c) topology control.

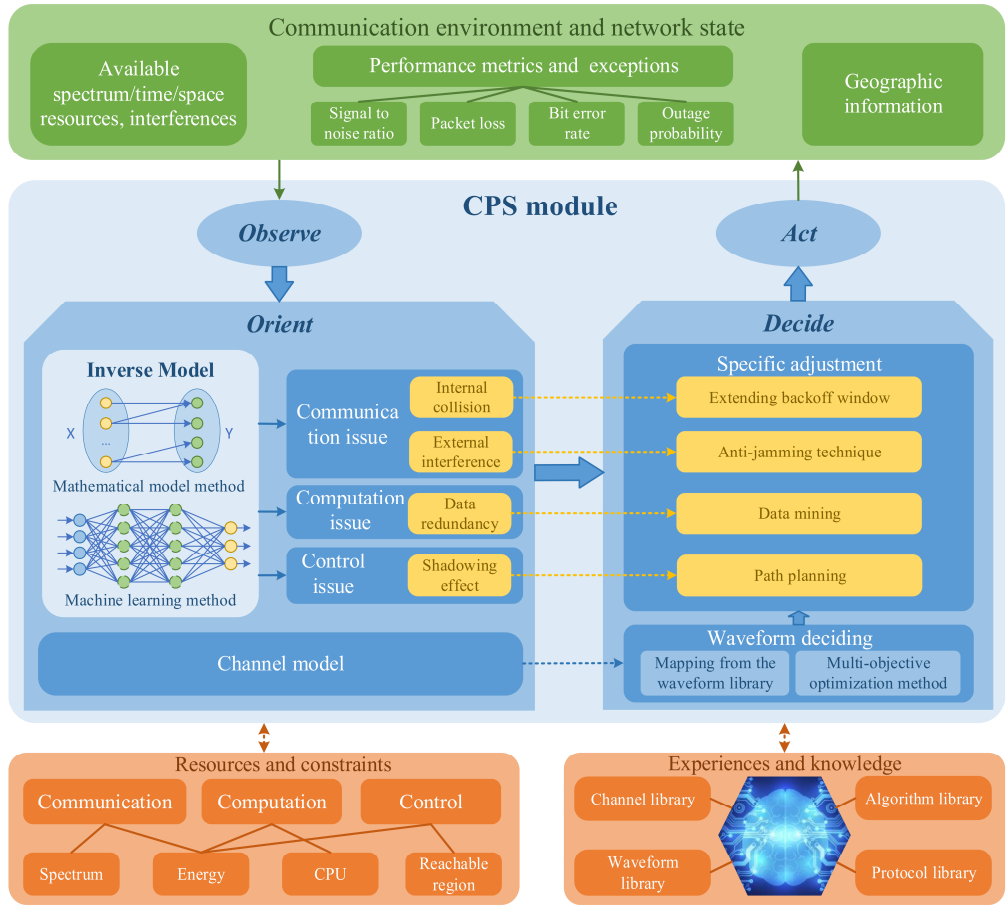

Figure 3. The CPS framework for UAV communications.

*iii) Topology control*: The scattered, unevenly distributed and possibly disconnected UAVs are aggregated to form a robust network with regular topology (e.g., triangular, ring and square).

## A CPS MODULE FOR UAV COMMUNICATIONS

Massive computation and control capabilities can be exploited to compensate UAV communications based on the coupling effects. In this section, a CPS module is proposed, which will instruct the coordination of various resources for better communication performance.

### THE FRAMEWORK OF CPS MODULE

A complete UAV communication process generally consists of four steps, including environment perception, environment understanding, decision making and final execution. It encompasses various elements, ranging from the physical to cyber domain. Different from the traditional communication, multidimensional decisions with respect to communication, computation and control, will be generated. Here, as shown in Fig. 3, we refer to the general OODA Loop (i.e., observe, orient, decide and act) to formulate the CPS module considering the commonalities with UAV communications.

**Observe:** UAVs will lay emphasis on perceiving the external environments, which include cyber and physical elements. The former focuses on the available spectrum/time/space resources and interferences. They can be obtained, for example, by adopting the simple energy detection algorithm. The latter mainly contains geographic information and reachable region. From some extent, the latter will determine the former, considering the cyber physical coupling effects in communication [4]. Besides, performance metrics and potential exceptions should also be monitored (e.g., SINR, packet loss, bit error rate and outage probability) since they will reflect the variations of external environments.

**Orient:** The explicit models, patterns and characteristics can be extracted from the raw data. For example, after obtaining the geographic information, UAVs can build the channel model by developing statistical propagation model [13]. As for jamming, its pattern (e.g., sweep or comb) can be recognized through interference detection algorithms. Further, by developing an inverse model, the issues can be oriented based on the extracted characteristics, performance metrics and exceptions. It means that UAVs could decide which kind of capability and resource should be invoked to facilitate communications. For example, when there exist external interferences, attacks or internal collisions, it can be classified as a communication issue since they can be handled through communication related methods. When UAVs encounter shadow fading channel, it should be oriented as a control issue because adjusting communication parameters can be unavailing and only movement may work. As the most important and difficult part, reasonable inverse rules should be established to orient the issues (i.e., communication, computation and control) and further deduce the causes (e.g., jamming, collision, attack, channel condition or data redundancy).

**Decide:** First of all, the communication waveform will be determined based on the channel characteristics. It can be matched from the off-the-shelf waveform library by building mapping mechanisms, or generated provisionally through multiobjective optimization methods. Then, UAVs will make some adjustments specific to the oriented problems in the former step from different layers. For example, for the communication problems, anti-jamming solutions can be carried out from the physical layer (e.g., error correcting codes and interference cancellation), or from the data link layer through employing multichannel MAC protocols. The transmission collision can be alleviated, for example, by extending the backoff window when DCF protocol is adopted. For the control problem, the trajectory of UAV will be planned with detailed heading, velocity and duration. For the computation problem, data mining techniques can be performed to lessen the date volume of each UAV. Multiagent decision technology can also be introduced to reduce the payload of the whole system. In many cases, the issues with respect to communication, computation and control can be jointly handled under a unified objective.

**Act:** UAVs will execute the decisions, for example, load the communication waveforms, fly to a certain location with scheduled speed and duration, or spend more computation resources to improve the communication precision.

The four steps constitute a closed loop. It enables continuous iterations and evolutions. Experiences and knowledges will be generated. They may include channel characteristics under various environments, mapping mechanisms from channels to waveforms and expanding waveform/algorithm/protocol libraries. All these will improve the operation efficiency of the CPS module, and hence the communication capability of UAVs in unknown and complex environments.

*THE INVERSE MODEL: TRADITIONAL METHOD VS. MACHINE LEARNING METHOD*

The communication performance metrics (i.e., effects) are input variables, and the precise causes are output variables. The mapping from causes to effects is often noninvertible. This happens when more than one cause results in the same effect. For example, external interferences, internal collisions, and channel quality deterioration all will lead to increased packet loss rate. This problem can be solved by expanding the dimensions of inputs. We can integrate effects from multiple levels and aspects, and introduce some cross-layer considerations. Therefore, the multidimensional inputs will contain the observations from physical domain (e.g., GIS and images) and the performance metrics from different network layers (e.g., RSSI, SINR, bit error rate, packet loss, throughput, delay and number of retransmissions). Two approaches can help to build this inverse model, i.e., the traditional method and machine learning method.

**Traditional method:** It is hard to derive the theoretical relationships between the performance metrics and causes. Nevertheless, the experience model can be rationally acquired. Generally, communication performance highly relies on the valid SINR. However, internal collisions, external spectrum interferences, ambient noise and shadowing effect all will result in the SINR reduction. Under these cases, the data cannot be correctly received and demodulated, which would deteriorate the bit error rate. Further, this will be transformed into repeating carrier sensing, backoff and retransmission in the MAC layer. Conversely, by combining the received signal of the physical layer with the protocol indicators of the MAC layer, we can deduce how the communication environment varies.

**Machine learning method:** Different from the traditional method pursuing a mathematical model, machine learning method requires a large number of data samples. The data can be collected through testing certain network performance metrics and variations under constructed communication environments. These performance metrics should be the attributes that can distinguish different communication environments. For example, the average number of retransmissions can be adopted to differentiate internal collisions from spectrum interferences and shadow fading. This is due to that under the latter two situations, the number of retransmissions is much likely to reach the maximum compared with the former. These data can be labeled to establish the inverse model based on supervised learning. For example, Bayesian learning is competent in deducing the causes according to the effects, that is, finding the maximum posterior probability. However, labeling the data is cumbersome. Therefore, unsupervised learning can be employed. It can exact the attributes autonomously and achieve anomaly detection and clustering among different communication environments. Besides, deep learning is also promising in establishing the inverse model.

## RESOURCES AND CONSTRAINTS IN UAV COMMUNICATIONS

Due to size, weight and power constraints of UAVs, various heterogeneous resources with respect to computation, communication and control, are restricted. Thus, they should be efficiently scheduled both to optimize their separate performance, and to optionally compensate communication for alleviating the bottleneck.

*HETEROGENEOUS RESOURCES AND CONSTRAINTS*

With the improvement of computation capabilities of UAVs, many AI algorithms are adopted (e.g., image recognition in reconnaissance and surveillance missions). They are resource hungry. Although offloading these computation-intensive tasks to resource rich infrastructures (e.g., the terrestrial base stations) seems to be promising, it appears at the expense of delays and communication resource consumptions. Therefore, the computation resources are not unconditionally abundant.

Compared with computation, communication resources are much more stringent. They are generally affected by external factors, including frequency spectrum and physical channel. First, due to the complex mission circumstances, UAVs usually suffer from spectrum scarcity and variations, and outside interferences. Second, considering the rough terrains where UAVs may be deployed, they often confront channel quality deterioration.

Flight control is dominated by missions and does not always serve for communication. Thus, communication should keep pace with the flight control, and solve the consequent troubles, including fluid topology, intermittent links and Doppler effects. For example, a UAV swarm may perform frequent formation transformation when tracking the randomly moving targets, however, it is not friendly to communication.

### ENERGY EFFICIENCY

As the common and most fundamental resource, the onboard energy is constrained by the limited payload capacity, which severely hampers UAV endurance. One direct solution is to timely replenish the onboard energy. For example, let the exhausting UAVs fly back for recharging. Yet, the accompanying drawbacks are apparent, including mission interruption and fluid topology. Exploiting inter-UAV cooperation to enable sequential energy replenishment is effective to deal with these problems. For example, only one UAV is arranged to leave the mission area for charging at one time interval, during which the service gap is temporarily undertaken by neighboring UAVs. However, these energy replenishment methods are not applicable to time-critical missions. Additionally, it may be time-consuming or dangerous to fly back to the remote charging station.

From another perspective, improving the energy efficiency may work. That is to efficiently exploit the limited energy to maximize the whole endurance. Since aircraft propulsion and wireless communication consume the most part of the energy, energy efficient operations can be carried out from two aspects. The first one is energy efficient flight. The movement of each UAV should be delicately planned by taking the energy consumption of each maneuver into consideration. Energy efficient flight can be integrated into path planning problem, where the velocity, acceleration and heading are jointly optimized. The other aspect is to achieve energy efficient communication. It aims at satisfying communication demands with minimum energy expenditure, for example, on signal processing and RF transmission. Moreover, it is verified that the propulsion power of rotary-wing UAVs firstly decreases and then increases as the velocity increases [14]. It means that hovering is not the most power conserving status. From this point, UAV can move to pursue high-quality links without consuming extra propulsion energy compared with hovering. This will bring about more successfully transmitted data and thus higher energy efficiency. Therefore, jointly planning communication and control is promising to achieve energy efficiency.

### ENERGY VALUE FUNCTION

For various scenarios, UAV communications are optimized with different metrics, for example, low latency in formation control scenario and high throughput in data collection scenario. Since energy is the common resource and also the constraint, its consumption can be optimized among communication, computation and control. Here, taking the communication planning problem of two rotary-wing UAVs under shadow effect as an example (as shown in Fig. 2a), we will illustrate how to build the energy value function for the joint optimization problem.

### COMPUTE-FLY-TRANSMIT PROTOCOL

Compute-fly-transmit is a very intuitional protocol. The communication process consists of three phases, including computing, flying and transmitting. For a UAV with given data block to transmit, it will first execute computation tasks, mainly including data preprocessing and decision making. The former tries to eliminate information redundancy. The latter targets for deciding waveform parameters. After the computation process, the UAV will fly to the destination and then perform data transmission. Note that, the UAV hovers during the computation and transmission periods. Only the sender moves if necessary, and the receiver keeps still. The whole communication process should be finished within a finite time horizon.

### ENERGY TRADEOFF

Computation resources are generally measured by the CPU frequency (in Hz) and occupation time (in seconds). Intuitively, the more the computation resources are designated to data preprocessing and decision making, the more the data redundancy will be eliminated and the better the waveform decisions are. Therefore, less data needs to be transmitted, and less propulsion and transmission energy will be consumed. This is reasonable since more computation resources enable more superior algorithms (e.g., AI algorithms), and thus more optimized decisions. Therefore, the propulsion and transmission energy reduction are generally at the cost of higher computation and hovering energy consumption. Besides, the UAV could travel much further to pursue a higher data rate. This strategy, though requiring additional energy for propulsion, reduces the time and hence energy for hovering and transmission. All these bring about a fundamental energy tradeoff among computing, flying and transmitting.

### ENERGY CONSUMPTION MODEL

**Computation energy consumption:** Given the CPU frequency, the duration allocated for the data preprocessing and waveform decision making processes should be respectively optimized.

It is assumed that there are certain bits of data to be transmitted. First, define the CPU occupation time assigned to data preprocessing operation. The acquired CPU cycles can be obtained. Further, we can calculate the eliminated redundant data bits, given the number of CPU cycles required to process one bit of redundant data. Therefore, the residual bits of data that need to be transmitted are determined. The energy consumption of data preprocessing can be obtained. It is determined by the CPU frequency, occupation time and a constant energy coefficient.

There is a gap between the achievable and theoretical channel capacity due to the employed waveform. First, define the CPU occupation time assigned to waveform decision making. The CPU cycles can be obtained. Assuming that more CPU cycles lead to more efficient waveform and hence a smaller gap, the gap coefficient can be formulated as a function of the assigned

| Simulation component | | Configuration |
|---|---|---|
| Simulated scenario | Distance between the two UAVs | 500 m |
| | Initial propagation condition | Shadow fading |
| Computation | CPU frequency | 1 GHz |
| | Computation energy coefficient | $10^{-28}$ |
| Control | UAV weight | 1.5 kg |
| | Propulsion power model | Depend on speed [14] |
| Communication | Radio frequency | 5 GHz |
| | Bandwidth | 2 MHz |
| | Noise power spectrum density | -169 dBm/Hz |
| | Maximum transmission power | 5 W |
| | Channel model | Probability-weighted LoS and NLoS |
| | Delay constraint | 25 s |

Table 1. Main simulation parameters.

CPU cycles. The function is monotonically decreasing and has a lower bound 1. Similarly, the energy consumption for making waveform decisions can be acquired. The total computation energy consumption and duration can also be calculated.

**Flying energy consumption:** It is assumed that the two UAVs are located at the same and fixed height. Considering the compute-fly-transmit protocol, we care more about the final location where the data transmission is performed. Therefore, the sender UAV can fly along a straight line to the final optimal location with a certain speed. The propulsion power model has been derived in [14], which depends on the speed. In order to minimize the propulsion energy, the sender UAV should fly with the *maximum-range speed* [14]. It can be obtained numerically from the propulsion power model. Therefore, according to the propulsion power under the maximum-range speed and the flying duration, the flying energy consumption can be obtained.

**Transmission energy consumption:** The sender UAV should head in the direction of the channel power gain being most improved. Due to shadow effect, it is not reasonable for the sender UAV to fly straight to the receiver UAV. Empirically, flying in the direction of deviating the initial line segment between them may work. By referring to the UAV-ground channel model in [15], we assume that the channel power gain is jointly determined by the distance and deviation angle. The deviation angle refers to the angle between the initial line segment and that during the flying. Given the flying duration and the maximum-range speed, the optimal flying direction can be numerically found by maximizing the channel power gain. Accordingly, the channel power gain at the final location can be obtained. Given transmission power, bandwidth, channel power gain and gap coefficient, the achievable data rate can be calculated according to Shannon's Law. Previously, we have obtained the residual bits of data that need to be transmitted. Therefore, the transmission duration and energy consumption can be acquired.

**Hovering energy consumption:** There are two hovering periods, including the computation phase and the transmission phase. The power consumption of hovering status can be obtained by substituting speed=0 into the propulsion power model. Thereby, the total hovering energy consumption can be acquired.

*ENERGY VALUE FUNCTION*

The energy value function can be calculated by adding the energy consumption of computation, flying, transmission and hovering, which needs to be minimized. To this end, the computation resources (including the CPU duration assigned to data preprocessing and waveform decision making), flying path (including the flying duration and direction) and transmission power should be jointly planned under the main constraint, that is, the total duration should not be more than a finite time.

*CASE SIMULATION AND DISCUSSION*

Based on the energy value function, we will evaluate the proposed CPS framework in improving the communication performance of UAVs. We take the data transmission of two UAVs under the shadowing effect as a case. The simulation setup is shown in Fig. 2a and Table 1. Initially, there exists a strong shadowing fading channel, which is incompetent in completing the transmission task within a delay constraint. It means that this cannot be simply oriented as a communication issue, but control and/or computation issues. Here, we compare two methods in dealing with this. One is the joint planning on communication and control (JP-CC) method, which is commonly adopted [12, 14]. The other is the proposed CPS method which capacitates the joint planning on communication, control and computation.

Figure 4 shows the total consumed energy versus the packet length. As expected, for both methods, the energy consumption increases with the increasing data packet. However, the consumed energy of the CPS method reduces by about 33 percent compared to the JP-CC method averagely. This is due to that the CPS method takes advantage of the computation process to eliminate the data redundancy and optimize the waveform. The data that needs to be transmitted can be reduced. All these contribute to the reduction of the total energy consumption.

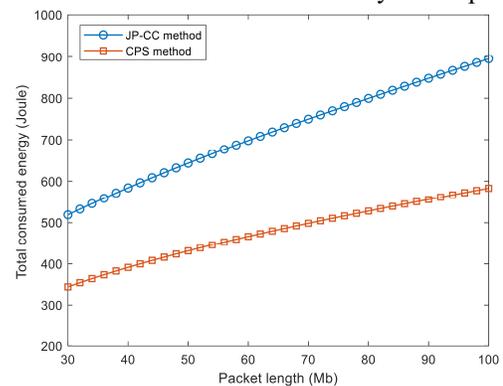

Figure 4. Total energy consumption.

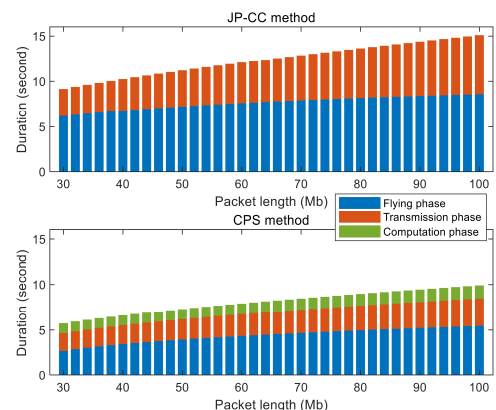

Figure 5. Communication delays.

Then, we will evaluate the delay of the whole communication process. Figure 5 depicts the respective and cumulative duration of the three phases for both methods. Similarly, the whole delay increases with the increasing data packet. However, the total elapsed time of the CPS method is much less than that of the JP-CC method. The magnitude of the decline reaches more than 30 percent. We can also see that the flying duration, in other words, the flying distance, is highly reduced after the CPS method is adopted. It implies that the UAV do not need to fly too far to complete the data transmission, which guarantees the global task and flying safety.

In conclusion, the proposed CPS method outperforms the JP-CC method in terms of the energy consumption and delay for UAV communications. Based on the improvement of flight control to communication, computation can further optimize the efficiency of communication and control. All these validate the effectiveness of the proposed CPS framework in dealing with UAV communication issues by empowering joint scheduling on communication, control and computation.

## Conclusion

In this article, we have proposed a novel CPS framework for UAV communications. It extends the dimension of communication decisions to computation and control through a cross-disciplinary method. The coupling effects of computation and control to communication are explicitly elaborated, which lays the foundation for the framework. Heterogeneous resources and design constraints are discussed. Furthermore, an energy value function is formulated for quantitatively instructing the protocol and algorithm design of UAV communications. Case simulations validate that the proposed CPS framework can highly reduce the energy consumption and communication delay, compared with the conventional joint planning on communication and control method. It is expected that CPS would push UAV communications step further to a more intelligent era.

## Acknowledgment

This work was supported by the National Natural Science Foundation of China under Grant 61931020, 61601480 and 61372099.

## Biographies

**Haijun Wang** (haijunwang14@nudt.edu.cn) received his B.S. degree from Shandong University, Jinan, China, in 2014, and M.S. degree from National University of Defense Technology (NUDT), Changsha, China, in 2016. He is currently pursuing the Ph.D. degree in College of Electronic Science and Technology, NUDT. His research interests include cognitive radio networks and UAV communications.

**Haitao Zhao** [SM'18] (haitaozhao@nudt.edu.cn) received his B.S., M.S. and Ph.D. degrees all from NUDT in 2002, 2004 and 2009 respectively. Presently, he is a professor with College of Electronic Science and Technology, NUDT. His research interests include cognitive radio networks and self-organized networks.

**Dongtang Ma** [SM'14] (dongtangma@nudt.edu.cn) received his B.S., M.S. and Ph.D. degrees all from NUDT in 1990, 1997, and 2004, respectively. Presently, he is a professor with the College of Electronic Science and Technology, NUDT. His research interests include physical layer security, intelligent communication and network.

**Jibo Wei** [M'04] (wjbhw@nudt.edu.cn) received his B.S. and M.S. degrees from NUDT in 1989 and 1992, respectively, and Ph.D. degree from Southeast University, Nanjing, China, in 1998. Presently, he is a professor with the College of Electronic Science and Technology, NUDT. His research interests include software defined radios, cognitive communications and networking, OFDM and MIMO, cooperative communications.